\documentclass[aps,prl,reprint,twocolumn,superscriptaddress,showpacs,amssymb]{revtex4-1}
\usepackage[english]{babel}
\usepackage{graphics}
\usepackage{graphicx}
\usepackage{epsfig}
\usepackage{amssymb}
\usepackage{dcolumn}
\usepackage{bm}
\usepackage{color}
\usepackage{natbib}
\usepackage{lineno}
\usepackage{sidecap}
\usepackage{verbatim}  
\usepackage{vector}  
\usepackage{wrapfig}
\usepackage{enumerate}
\usepackage{sidecap}
\usepackage{amsmath}
\usepackage{hyperref}

\begin{document}

\title{Unusual Dirac fermions on the surface of noncentrosymmetric $\alpha$ - BiPd superconductor}

\author{S. Thirupathaiah}
\email{t.setti@sscu.iisc.ernet.in}
\affiliation{Solid State and Structural Chemistry Unit, Indian Institute of Science, Bangalore, Karnataka, 560012, India.}
\author{Soumi Ghosh}
\affiliation{Department of Physics, Indian Institute of Science, Bangalore, Karnataka, 560012, India.}
\author{Rajveer Jha}
\affiliation{CSIR-National Physical Laboratory, New Delhi 110012, India.}
\author{E. D. L. Rienks}
\affiliation{Leibniz Institut f\"ur Festkörper- und Werkstoffforschung IFW Dresden, D-01171 Dresden, Germany.}
\author{Kapildeb Dolui }
\affiliation{Department of Physics, Indian Institute of Science, Bangalore, Karnataka, 560012, India.}
\author{V. V. Ravi Kishore}
\affiliation{Solid State and Structural Chemistry Unit, Indian Institute of Science, Bangalore, Karnataka, 560012, India.}
\author{B. B\"uchner}
\affiliation{Leibniz Institut f\"ur Festkörper- und Werkstoffforschung IFW Dresden, D-01171 Dresden, Germany.}
\author{Tanmoy Das}
\affiliation{Department of Physics, Indian Institute of Science, Bangalore, Karnataka, 560012, India.}
\author{V. P. S. Awana}
\affiliation{CSIR-National Physical Laboratory, New Delhi 110012, India.}
\author{D. D. Sarma}
\affiliation{Solid State and Structural Chemistry Unit, Indian Institute of Science, Bangalore, Karnataka, 560012, India.}
\author{J. Fink}
\affiliation{Leibniz Institut f\"ur Festkörper- und Werkstoffforschung IFW Dresden, D-01171 Dresden, Germany.}
\date{\today}

\begin{abstract}
     Combining multiple emergent correlated properties such as superconductivity and magnetism within the topological matrix can have exceptional consequences in garnering new and exotic physics. Here, we study the topological surface states from a noncentrosymmetric $\alpha$-BiPd superconductor by employing angle-resolved photoemission spectroscopy (ARPES) and first principle calculations. We observe that the Dirac surface states of this system have several interesting and unusual properties, compared to other topological surface states. The surface state is strongly anisotropic and the in-plane Fermi velocity varies rigorously on rotating the crystal about the $y$-axis. Moreover, it acquires an unusual band gap as a function of $k_y$, possibly due to hybridization with bulk bands, detected upon varying the excitation energy. Coexistence of all the functional properties, in addition to the unusual surface state characteristics make this an interesting material.
\end{abstract}

\maketitle


Superconductivity in the presence of spin-orbit coupling, magnetism, and topological surface states can have exceptional consequences, such as chiral pairing~\cite{Nandkishore2012}, coexistence of singlet and triplet pairings~\cite{Bauer2004, Frigeri2004}, Majorana bound state~\cite{Sato2009a}, topological superconductivity~\cite{Chadov2010a, Lin2010a} and also emergent supersymmetry~\cite{Grover2014}. Such pairing is long investigated in several noncentrosymmetric heavy-fermion superconductors~\cite{Bauer2004, Frigeri2004}, but without a topological state. Furthermore, tremendous efforts are paid to engineer topological superconductivity via growing heterostructures of topological insulators and superconductors~\cite{Beenakker2013}.



$\alpha$-BiPd is recently synthesized with all the aforementioned properties obtained intrinsically, and thus provides the long-sought material for new experiments and applications. So far, there have been few works on this material, mainly studying the magnetic and transport phenomena~\cite{Joshi2011, Mondal2012, Matano2013}. Again, for the spectroscopic investigations, there is only one ARPES~\cite{Neupane2015} and one scanning tunnelling spectroscopy (STS) study~\cite{Sun2015}. The ARPES work reported the electronic structure of this compound with the detection of topological Dirac cone at -0.7 eV below the Fermi level ($E_F$), without detailing the properties of Dirac cone, while the STS work reported the states above the Fermi level.


In this paper we report several interesting and unusual properties of the topological Dirac states present on the surface of this noncentrosymmetric $\alpha$-BiPd superconductor by employing ARPES and first principle calculations. We detect the surface states that are having a Dirac node at a binding energy of 0.7 eV below Fermi level ($E_F$) at the $\Gamma$ point dispersing along the $\Gamma-X$ high symmetry line. Upon varying the photon energy,  we notice surface states that are gapped as a function of $k_y$ at the node due to a possible hybridization with bulk bands, a unique feature of the surface state that is not disclosed in this compound so far. Upon varying the photon polarization we identify the orbital character of the detected bands.


\begin{figure*}
	\centering
		\includegraphics[width=0.85\textwidth]{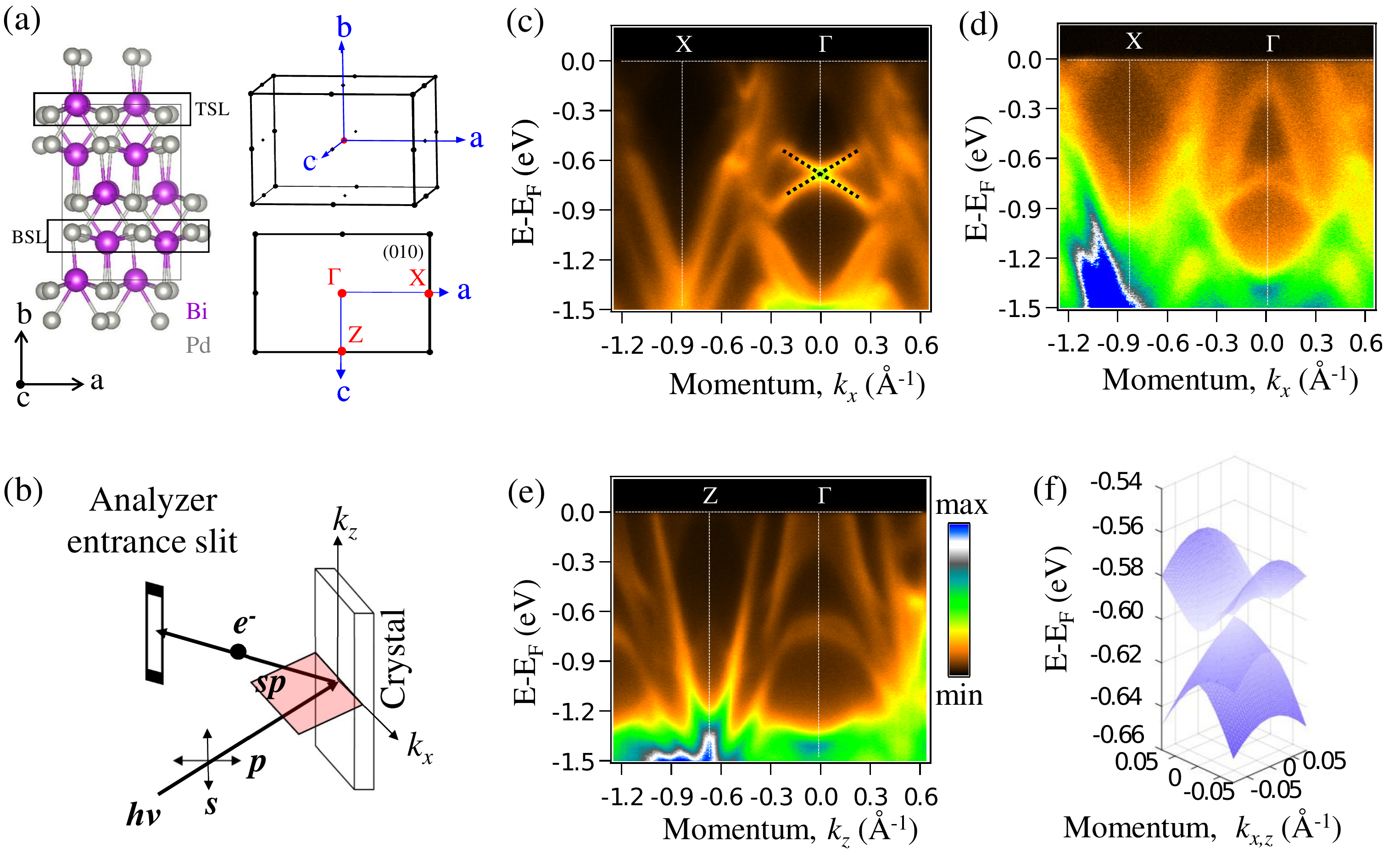}
	\caption{(Color online)(a) Crystal structure of $\alpha$-BiPd (left), 3D Brillouin zone (top right) and 2D Brillouin zone (bottom right) projected onto the (010) plane in which we locate the respective positions of the high symmetry points. Top surface layer (TSL) and bottom surface layer (BSL) are shown on the crystal structure. Preferable cleavage plane of the crystal is between any two Bi layers. The lattice parameters are given by $a$= 7.33 \AA,  $b$= 10.86 \AA ~and  $c$ = 8.82 \AA. In (b) we schematically show a typical measuring geometry in which  $s$- and $p$-plane polarized lights are defined with respect to the analyzer entrance slit and the scattering plane (SP). Note here that $k_x$ and $k_z$ represent in-plane momentum vectors, while $k_y$ is the out-of-plane momentum vector. (c) and (d) show energy distribution maps (EDMs) measured using $p$ and $s$ polarized lights, respectively along the $\Gamma-X$  high symmetry line with a photon energy of 75 eV. Similarly, (e) shows the EDM measured using $p$ polarized light along the $\Gamma-Z$ high symmetry line with 75 eV photon energy. (f) shows a 3D view of the Dirac surface state from DFT calculations along $k_x$ and $k_z$.}
	\label{1}
\end{figure*}

We further show that the Dirac fermions in $\alpha$-BiPd are highly anisotropic on rotating the crystal about the $y$-axis such that, in going from $\Gamma-X$ to $\Gamma-Z$  the massless linear dispersive Dirac states become flat dispersive massive fermions. Therefore, the Dirac states found in this compound are of one dimensional (1D) character which could provide a natural route for the quantum wires. In general, the Dirac states found on the surface of 2D compounds such as graphene~\cite{Novoselov2005} and topological insulators~\cite{Hasan2010},  in a few Bi related compounds (Na$_3$Bi and Bi)~\cite{Wang2012a} and other 3D compounds like,  Cd$_3$As$_2$~\cite{Neupane2014} and PbSnSe~\cite{Liang2013} are characterized to be nearly isotropic and forming a Dirac cone with isotropic Fermi velocity and effective mass of the fermions. On the other hand, there are theoretical predictions~\cite{Zhang2011b, Virot2011,Yan2012} suggesting anisotropic Dirac states with massless fermions in one direction and massive fermions on the other direction. Till date, only a few experimental papers are reporting the asymmetric Dirac states in Ru$_2$Sn$_3$~\cite{Gibson2014},  Sr(Ca)MnBi$_2$~\cite{Feng2014}, and $\beta$-Bi$_4$I$_4$~\cite{Autes2015}.  Unusual band structure of this material, surface state is gapped and asymmetric, make this a unique compound which will attract tremendous future research interests.

Single crystals of stoichiometric $\alpha$-BiPd were grown by a self-flux melt growth method. The crystals have a platelet-like shape with shiny surfaces. As grown crystals show a superconducting transition at $T_c$ $\approx$ 4 K.  Further elemental analysis on these single crystals are reported elsewhere~\cite{Jha2016}. ARPES measurements were carried out in BESSY II (Helmholtz Zentrum Berlin) synchrotron radiation facility at the UE112-PGM2b beam line using the "1$^2$-ARPES" end station equipped with SCIENTA R8000 analyzer. The total energy resolution was set between 15 and 35 meV, depending on the applied photon energy. Samples were cleaved $\textit{in situ}$ at a sample temperature of  50 K. All measurements were carried at a sample temperature $T\approx$ 50 K.

Band structure calculations are performed on the orthorhombic crystal structure of $\alpha$-BiPd~\cite{Bhatt1979}, though the primitive unit cell is of monoclinic~\cite{Joshi2011}, using density functional theory within the generalized gradient approximation (GGA) of Perdew, Burke and Ernzerhof (PBE) exchange and correlation potential~\cite{Perdew1996} as implemented in the Vienna \textit{ab-initio} simulation package~\cite{Kresse1996}. Projected augmented-wave (PAW) ~\cite{Kresse1999}~ pseudo-potentials are used to describe core electrons and spin-orbit coupling as included in the calculations.The electronic wavefunction is expanded using plane waves up to a cutoff energy of 500 eV. Brillouin zone sampling is done by using a (6$\times$4$\times$5) Monkhorst-Pack $k$-grid. Both atomic positions and cell parameters are allowed to relax, until the forces on each atom are less than 0.01 eV/\AA. In order to simulate surface effects, we used 1$\times$4$\times$1 super-cell for the (010) surface, with vacuum thickness larger than 15 \AA. Our slab calculations are in good agreement with prior calculations~\cite{Neupane2015}. It is observed that the surface states calculated on an orthorhombic crystal match better with the experiment than calculated on a monoclinic crystal, which yields the Dirac states at the Brillouin zone edge~\cite{Sun2015} instead of at the center (see below).


\begin{figure*}
	\centering
		\includegraphics[width=0.85\textwidth]{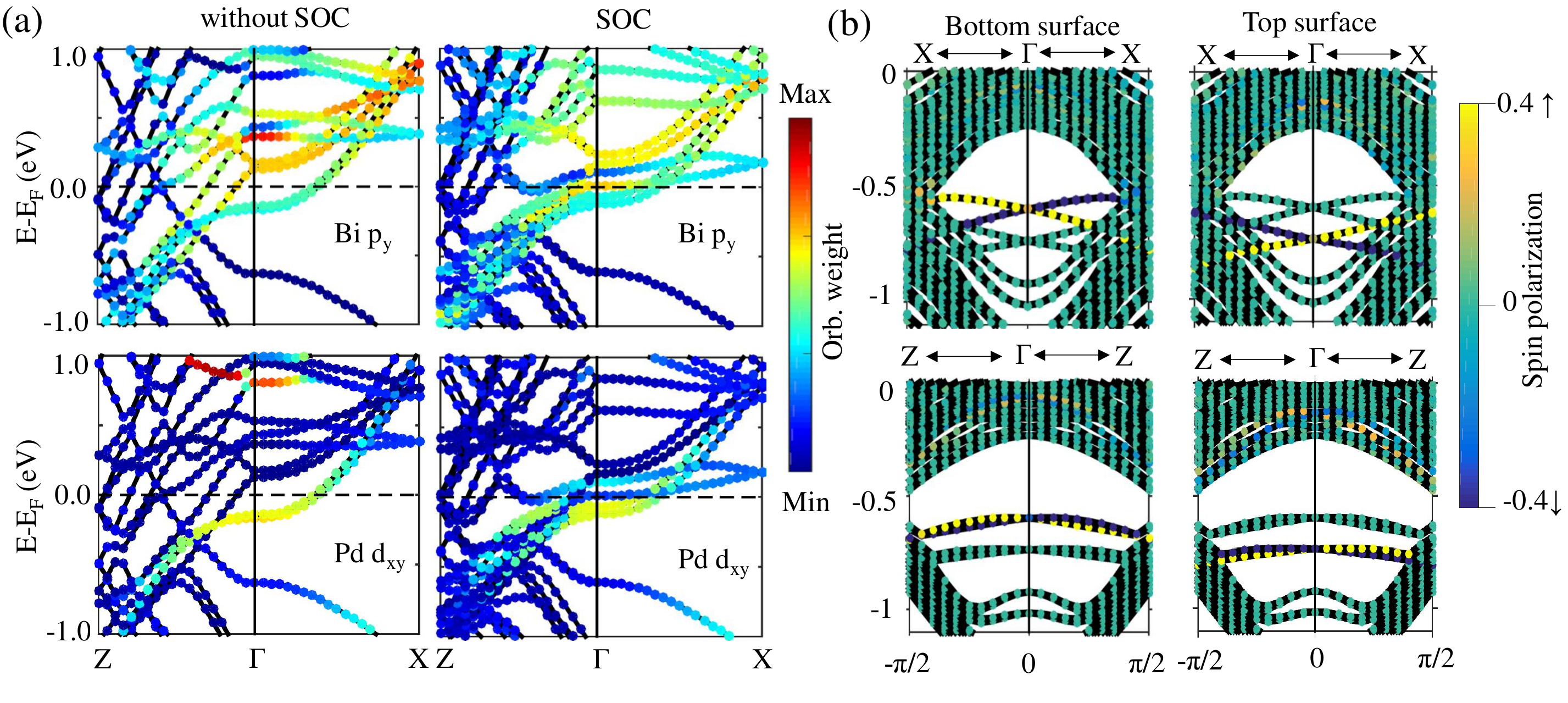}
	\caption{(Color online) DFT-GGA band structure calculations. (a) Bulk band structure without the spin-orbit coupling (SOC) for Bi $p_y$ orbital character (top left) and for Pd $d_{xy}$ (bottom left). Similarly, the bulk band structure with spin-orbit coupling is shown for $p_y$ (top right) and for Pd $d_{xy}$ (bottom right). (b) Slab calculations for the surface band structure along the high symmetry lines $\Gamma-X$ and $\Gamma-Z$ from the top and bottom surface layers with SOC for Bi $p_y$ and Pd $d_{xy}$ orbitals.}
	\label{2}
\end{figure*}

ARPES data of $\alpha$-BiPd are shown in Figs.~\ref{1}(c)-(e). From the energy distribution map (EDM) shown in Fig.~\ref{1}(c), we can clearly see that the Dirac states at $\Gamma$ are linearly dispersing  (black dashed lines) along the $\Gamma-X$ high symmetry line with a Dirac node at a binding energy of -0.7 eV. From the EDMs shown in Figs.~\ref{1}(c) and ~\ref{1}(e) we can further see that as we rotate the crystal about $y$-axis from $\Gamma-X$ to $\Gamma-Z$ direction, eventually the linearly dispersive surface states become almost flat,  such that the fermions become increasingly massive as we go from $\Gamma-X$ to $\Gamma-Z$ while still preserving the same spin polarization for the bands [see Fig.~\ref{2}(b)] in both directions. From this observation it is clear that the Dirac states in this compound are highly one dimensional (1D). This is further demonstrated in a 3D view of the calculated band structure shown in Fig.\ref{1}(f) and using the ARPES data shown in Fig. S2 (supplementary material). This observation is in good agreement with a theoretical prediction on HgS, which suggests that such an asymmetric Dirac state exists when there is a broken fourfold rotation symmetry~\cite{Virot2011}.  In addition, there exist very few experimental reports discussing the asymmetric Dirac cone~\cite{Feng2014, Gibson2014, Autes2015}. The states reported in Ref.~\onlinecite{Feng2014} for the case of Sr/CaMnBi$_2$ cannot be classified into one of the 1D, 2D or 3D type of Dirac cone because, it is in the arc shape rather than cone shape and moreover the Fermi velocity varies between  $\Gamma-M$ and $\Gamma-M'$. The Dirac states in Ru$_2$Sn$_3$ are of quasi-1D~\cite{Gibson2014}, while the states in  $\beta$-Bi$_4$I$_4$ are of 1D but dispersing in the out-of-plane~\cite{Autes2015} in contrast to the in-plane dispersing asymmetric Dirac states of $\alpha$-BiPd.  Thus, the anisotropic Dirac states of this compound are different compared to the existing asymmetric Dirac states.


 Next, to delineate if the Dirac states seen in Fig.~\ref{1}(c) arises from a trivial Rashba-type spin-orbit coupling (SOC) or from the topological nature of the bulk band, we first study the band inversion property without and with SOC. Fig.~\ref{2}(a) compares the band structure without SOC (left panel) and with SOC (right panel), overlayed with the relevant orbital weights for Bi $p_y$ (top) and Pd $d_{xy}$ (bottom). We observe a clear band inversion with SOC [Fig.~\ref{2}(a)] at the $\Gamma$-point in which the Bi $p_y$ orbital drops below the Fermi level, while some Pd $d_{xy}$ orbital character is shared to the conduction bands. Since, this is the only time-reversal invariant k-point, where we detect a band inversion, the system represents a non-trivial topological metal~\cite{Fu2007}. Interestingly,  we further notice that the band inversion in $\alpha$-BiPd  is of $pd-pd$ type whereas the band inversion reported so far in the topological systems are of $s-p$~\cite{Hasan2010, Qi2011} and $p-d$ in Ru$_2$Sn$_3$~\cite{Gibson2014}.

 %


On comparing the EDMs shown in Figs.~\ref{1} with the bulk [Fig.~\ref{2}(a)] and surface [Fig.~\ref{2}(b)] band structures from the calculations, we infer that the bands near the Fermi level in Figs.~\ref{1}(c), (d) and (e) are mostly from the bulk. Further, from the slab calculations shown in Fig.~\ref{2}(b) we could see that at $\Gamma$ the Dirac states present inside the bulk gap and adiabatically connect the bulk bands, which is a requirement for the topological state~\cite{Hasan2010, Qi2011}. From the calculations,  two Dirac nodes have been identified at -0.6 eV and -0.73 eV from the bottom and top surfaces layers, respectively. Degeneracy of the surface states between top and bottom layers is lifted by noncentrosymmetry of the crystal which induces the electric field gradient intrinsically along the $b$-axis. However, from our ARPES data [Fig.~\ref{1}(c))] we could resolve only one Dirac node at -0.7 eV, which is supposed to be from the top surface layer according to our calculations. This observation of the Dirac node at -0.7 eV is consistent with Ref.~\cite{Neupane2015}. With the help of our first principle calculations we ascribe Bi $p_y$ and Pd $d_{xy}$ orbital characters to the Dirac states. This is further confirmed by our polarization dependent ARPES measurements: $p$ polarized light detects the bands predominantly contributed by the even parity states,  Bi $p_y$ and Pd $d_{xy}$, according to the measuring geometry shown in Fig.~\ref{1}(b). Therefore, $s$ polarized light cannot detect the bands composed of these orbital characters. This could explain why the spectral weight of the surface states is higher when probed with $p$ polarized light [see Fig.~\ref{1}(c)] compared to $s$ [see Fig.~\ref{1}(d)].


\begin{figure}
	\centering
		\includegraphics[width=0.45\textwidth]{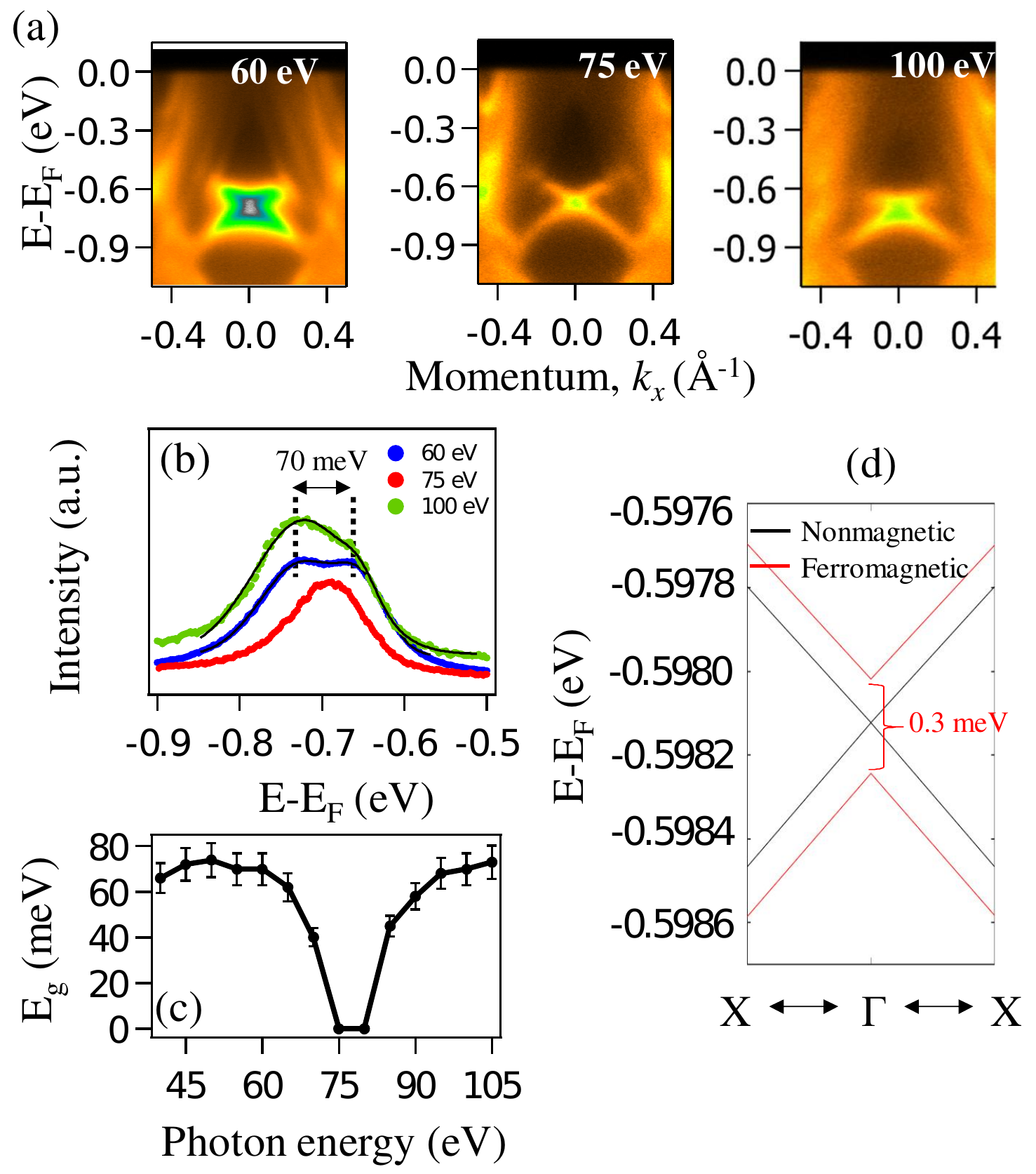}
	\caption{(Color online) Top panels in (a) show representative photon energy dependent EDMs taken along the $\Gamma-X$ direction with 60, 75 and 100 eV measured using $p$-polarized light. (b) Depicts energy distribution curves (EDCs) taken at $k_\parallel$=0 from the EDMs shown in (a). In (b) black curve is a fit to the EDCs using double Lorentzian functions. (c) Depicts the experimentally obtained photon energy dependent band gaps ($E_g$) at the Dirac node. (d) Depicts the DFT band structure calculated without magnetic order (black color) and with the imposed magnetic moments on the surface Pd atoms(red color), in which we could notice a band gap ($E_g$) opening at the Dirac node in the presence of magnetic order.}
	\label{3}
\end{figure}



Next, in Fig.~\ref{3}(a) we show representative photon energy dependent EDMs measured using $p$ polarized light along $\Gamma-X$, with photon energies 60, 75 and 100 eV. The EDMs measured with the other photon energies can be found in Fig. S3 (supplementary material).  In Fig.~\ref{3}(b) we show energy dispersive curves (EDCs) taken at $k_\parallel$=0 from the EDMs shown in ~\ref{3}(a). We have extracted the band gaps using a fit with double Lorentzian functions to the EDCs from all photon energies. In Fig.~\ref{3}(b) the black curves represent two of such EDC fittings from 60 and 100 eV.  Fig.~\ref{3}(c) shows the gap ($E_g$) as a function of photon energy. From Fig.~\ref{3}(c) we can clearly see that the gap increases sharply when detected with the photon energies below 75 eV or above 80 eV and a maximum gap of 70 meV is recorded with photon energies 60 eV and 100 eV [see Fig.~\ref{3}(c)].


There are multiple ways a band gap can open at the Dirac node~\cite{Liu2009i, Biswas2010, Sanchez-Barriga2016}.  Without breaking the time reversal symmetry, a band gap can open in a finite size system where the top and bottom surface states can hybridize~\cite{Das2013}, or due to hybridization between surface and bulk states if they lie in the same energy scale or with breaking time reversal symmetry due to magnetism~\cite{Liu2009i, Biswas2010}. In our system, the first scenario is ruled out. Interestingly, a small ferromagnetic order has been recently detected in this compound~\cite{Jha2016}.  Within first principle calculations, we also find a small out-of-plane FM order coming from the Pd atom,  with magnetism of the order of 0.01 $\mu_B$. Using this method, we obtained a band gap of 0.3 meV at the node [see Fig.\ref{3}(d)] which is almost negligible compared to the experimental band gap (70 meV), suggesting that the week ferromagnetism could not cause such a high band gap. After ruling out the above possible scenarios for this effect, we would propose that there can be hybridization between surface and bulk states which might lead to a gap at the node, since the surface state is buried much below the Fermi level and there are bulk bands present at the Dirac point.  Therefore, we suggest that the photon energies 75 eV and 80 eV are probing the surface bands in which we see gapless Dirac states, while all other photon energies probing from deeper and thus detecting the surface bands that are hybridized with bulk. This argument is consistent with the earlier papers reporting that a topological surface state has a substantial penetration depth in $\beta$-Bi$_4$I$_4$~\cite{Autes2015} and Bi$_2$Se$_3$~\cite{Li2009j}, but it is yet to be clear to us how sensitive are the surface states of $\alpha$-BiPd to a change in photon energy as small as 5 eV. On the other hand, a possible $k_y$ dispersion of the bulk band crossing $\Gamma$ at -0.7 eV [see Fig.~\ref{2}(a)] can also lead to a photon energy dependent gap. In this case, in Fig.~\ref{3}(c), one should find at least one more photon energy ($k_y$-point) where the gap closes either in the range of 40-70 eV or 85-105 eV as we probe more than 3 Brllouin zones, calculated with the Bismuth inner potential of 20 eV,  along $k_y$ direction. But this is not the case as seen in Fig.~\ref{3}(c). Since currently we do not have a clear interpretation on the observation of photon energy dependent band gap, we suggest for further experimental and theoretical studies to fully understand this very interesting observation.

In summary, we have studied the band structure of $\alpha$-BiPd noncentrosymmetric superconductor using ARPES and first principle calculations. Dirac surface states are detected with a node at -0.7 eV from our studies. We notice that these Dirac states are highly anisotropic, which means along $\Gamma-X$ symmetry line these show linear band dispersion and along the $\Gamma-Z$ these states barely disperse. We have disentangled the orbital character of the surface states using polarization dependent measurements.  Interestingly, we detect a photon energy dependent band gap at the Dirac node for the surface states due to a possible hybridization with bulk bands. Since these Dirac states are in close proximity to a bulk superconductor, we think, the interplay between the topological surface states and the bulk superconductivity can provide a platform for new quantum information such as chiral superconductivity~\cite{Nandkishore2012} or Majorana zero modes~\cite{Sato2009a} in these compounds. Further experimental and theoretical studies are required to fully understand the photon energy dependent band gap and investigate the exotic implications of these findings.




T.S. acknowledges support by the Department of Science and Technology (DST) through INSPIRE-Faculty program (Grant number: IFA14 PH-86). T.S. acknowledges greatly the travel support given by IFW Dresden for the measurements. V.P.S.A and R.V. acknowledge the financial support from the Govt. of India through the DAE-SRC outstanding researcher award scheme.
\bibliography{BiPd}

\end{document}